\documentclass[fleqn,usenatbib]{mnras}

\usepackage{todonotes}
\usepackage[T1]{fontenc}

\DeclareRobustCommand{\VAN}[3]{#2}
\let\VANthebibliography\thebibliography
\def\thebibliography{\DeclareRobustCommand{\VAN}[3]{##3}\VANthebibliography}

\usepackage{graphicx}	
\usepackage{amsmath}	
\usepackage{amssymb}	
\usepackage{subfigure}
\usepackage{orcidlink}
\usepackage{newtxtext,newtxmath}

\title[Dependence of peculiar velocity on stellar mass]{Dependence of peculiar velocity on the host properties of the gravitational wave sources and its impact on the measurement of Hubble constant}

\author[Nimonkar and Mukherjee]{Harshank Nimonkar$^{1}$\thanks{harshank.nimonkar@xaviers.edu.in; nimonkar.harshank@gmail.com}\orcidlink{0009-0001-8290-119X},
Suvodip Mukherjee$^{2}$\thanks{suvodip@tifr.res.in}\orcidlink{0000-0002-3373-5236}\\
$^{1}$Department of Physics, St. Xavier's College, 5, Mahapalika Marg, Mumbai 400001, India\\
$^{2}$ Department of Astronomy \& Astrophysics, Tata Institute of Fundamental Research, 1, Homi Bhabha Road, Colaba, Mumbai 400005, India
}

\date{Accepted XXX. Received YYY; in original form ZZZ}

\pubyear{2023}

\begin{document}
\label{firstpage}
\pagerange{\pageref{firstpage}--\pageref{lastpage}}
\maketitle

\begin{abstract}
Accurate measurement of the Hubble constant from standard sirens such as the gravitational wave (GW) sources with electromagnetic counterparts relies on the robust peculiar velocity correction of the redshift of the host galaxy. We show in this work that the peculiar velocity of the host galaxies exhibits a correlation with the properties of the host galaxy primarily such as its stellar mass and this correlation also evolves with redshift. As the galaxies of higher stellar mass tend to form in galaxies with higher halo masses which are located in spatial regions having a non-linear fluctuation in the density field of the matter distribution, the root mean square (RMS) peculiar velocity of more massive galaxies is higher. As a result, depending on the formation channel of the binary compact objects, the peculiar velocity contamination to the galaxies will be different. The variation in the peculiar velocity of the host galaxies can lead to a significant variation in the estimation of the Hubble constant inferred using sources such as Binary Neutron Stars (BNSs). For the network of GW detectors such as LIGO-Virgo-KAGRA (LVK), LVK+LIGO-India, and Cosmic Explorer+Einstein Telescope, the variation in the precision of Hubble constant inferred from 10 bright siren events can vary from $\sim 5.4 - 6\%$, $\sim 4.5 - 5.3\%$ and $\sim 1.1 - 2.7\%$ respectively. The impact of such a correlation between peculiar velocity and stellar mass on the inference of the Hubble constant is not only limited to GW sources but also applicable to type-Ia supernovae.
\end{abstract}



\section{Introduction:}
Gravitational wave (GW) sources such as binary neutron star (BNS), binary black hole (BBH), and neutron star-black hole (NSBH) systems \citep{abbott_2016,GW170817, Abbott_2021} are called standard sirens as they offer a unique advantage by providing an independent and accurate measure of their luminosity distance to the source \citep{schutz}. However, due to the nature of GW observations, there is an inherent degeneracy between the luminosity distance and the inclination angle, which is the orientation of the orbital plane with respect to the observer. The degeneracy between these two parameters presents a challenge in accurately determining the astrophysical properties of the sources solely based on GW observations unless the distance-inclination angle degeneracy can be lifted by measuring the two polarizations of GW, using higher order modes for asymmetric mass systems, or improving the inference of inclination angle using observations from electromagnetic (EM) bands \citep{Holz_2005, Nissanke_2010, Nissanke_2013, Mooley_2018a, Hotokezaka_2018, Mukherjee_2020measurement}. 

Bright standard sirens having an observable EM counterpart, allow for an independent measurement of the redshift of the host galaxy which can be used along with the measured luminosity distance to measure cosmological parameters that impact the expansion history of the Universe such as the Hubble constant ($H_0$) \citep{schutz, Holz_2005, Nissanke_2010, Nissanke_2013, Mooley_2018a, Hotokezaka_2018,Chen_2018, Feeney:2018mkj, Mortlock:2018azx}. Such measurements are performed for the event GW170817 by combining the measurement of luminosity distance and redshift \citep{Abbott_2017b}. A further improvement in the measurement of $H_0$ was possible by including the constraints on inclination angle from jet measurement \citep{Hotokezaka_2018}. This measurement was further revised by including a better measurement of peculiar velocity \citep{Mukherjee_2020measurement,Howlett_Davis_2020, Nicolau_2020}. Another GW event GW190521 \citep{LIGOScientific:2020iuh} with tentative EM association made it possible to obtain a weak measurement on H$_0$ \citep{Mukherjee_2021, Gayathri:2020mra, Chen:2020gek}. 

Along with bright siren measurement, measurements from dark standard sirens made using statistical-host identification technique \citep{Soares-Santos_2019, Fishbach_2019, Abbott_2021, LIGOScientific:2021aug, Palmese_2023, Gair:2022zsa}, mass distribution \citep{Mastrogiovanni_2021, Leyde:2022orh, Karathanasis:2022rtr} and more recently by using the spatial clustering of the GW sources with galaxies \citep{Mukherjee_2022} using cross-correlation method \citep{PhysRevD.93.083511, Mukherjee:2019wcg,Mukherjee:2020hyn, Diaz:2021pem} have emerged as an independent probe having the potential to reach the required percent level precision on $H_0$, thus advancing towards resolving the well established Hubble tension \citep{Hinshaw_2013,Planck_2014,Planck_2016,wong_2019,Planck_2021,Riess_2022,Abdalla:2022yfr}.

In an isotropically expanding universe \citep{Lemaitre:1927,1929PNAS...15..168H,Lemaitre:1931}, the gravitational instability within large-scale structures leads to deviations from the smooth Hubble flow, introducing irregularities in the motion of the host galaxies of GW sources. Consequently, these galaxies acquire an additional velocity component beyond their recessional velocity (due to the Hubble flow). The additional velocity component of the host galaxy of a GW source, called peculiar velocity, contaminates the redshift measurements from the EM counterpart. The contribution from the peculiar velocity (few hundreds of km s$^{-1}$) to the motion of a GW host is significant up to a redshift of $z=0.05$ ($\approx$15,000 km s$^{-1}$). The contamination from individual sources degrades the accuracy and precision in the measurement of $H_0$ if not accounted for appropriately \citep{Mukherjee_2020measurement,Howlett_Davis_2020, Nicolau_2020}.

As we show in the following sections, the peculiar velocity dispersion depends on the property of the galaxy such as the stellar mass. The galaxies with higher stellar mass tend to have a larger peculiar velocity dispersion than the sources with lower stellar mass. As a result, depending on the formation of the GW sources in galaxies with higher or lower stellar mass, they will have different amounts of contamination.  With the current sensitivities of the ground-based GW detectors, we expect confident bright siren detection up to $\sim$ 200 Mpc \citep{Abbott_2020} which makes it a matter of concern to understand the role of their host properties and the possible contamination from the peculiar velocity that can impact the estimation of $H_0$.

The paper is structured as follows, in section \ref{section:motivation}, we discuss the motivation behind this work, followed by a brief review of the peculiar velocity formalism employed in this work in section \ref{section:vp model}. In section \ref{section:GW source population modelling}, we discuss our approach toward modeling a population of bright sirens. In section \ref{section:peculiar velocity and hubble constant}, we simulate bright siren events in these populations of galaxies considering different ground-based GW detector configurations followed by a Bayesian parameter estimation of the luminosity distance and inclination angle of the simulated GW sources. Finally in section \ref{section:conclusion} we summarize the work and discuss future outlook.

\section{Motivation} \label{section:motivation}
The presence of perturbations in the matter density leads to physical motion of the galaxies with respect to the cosmic rest frame, known as the peculiar velocity. The study of peculiar velocities has implications for our understanding of dark matter and dark energy and for interpreting observed cosmological phenomena such as redshift space distortions \citep{Fry_1994,Song_2009,Hudson_2012,Zheng_2013,Kim_2020,Cuesta-Lazaro_2020,Adams_2020,Turner_2022}.

The total velocity of the host galaxy of a GW source is the sum of the velocity due to Hubble flow, the peculiar velocity, and the velocity of the observer. Thus, in the expansion frame of reference, the peculiar velocity of the host galaxy $v_p$ is related to the velocity of the source $\vec v_s$ and that of the observer $\vec v_o$ as 
\begin{equation}
    v_p=(\vec v_s-\vec v_o).\hat{n}
\end{equation}
where $v_p$ is directed along the line of sight denoted by $\hat{n}$. A positive value of $v_p$ implies that the host galaxy is moving away from the observer. The difference in velocities of the source and the observer leads to a difference in the observed and true (cosmological) redshift as,
\begin{equation} \label{vp contamination eqn}
    \left(1+z_{\rm obs}\right)=(1+z_{\rm true})(1+\frac{v_p}{c}).
\end{equation}

For a GW source at a cosmological redshift $z_{\text{true}}$, its luminosity distance is given via the distance-redshift relation as,
\begin{equation} \label{eq:dL-z}
    d_L = \frac{c(1+z_{\rm true})}{H_0} \int_{0}^{z_{\rm true}}\frac{dz}{E(z)},
\end{equation} 
where the term $E(z)= \sqrt{\Omega_m(1+z)^3+ (1-\Omega_m)}$ is the ratio of the Hubble parameter ($H(z)$) to the Hubble constant $H_0$, written in terms of matter density $\Omega_m$ for a flat $\Lambda$CDM cosmological model.  However, for low redshifts, $H(z)$ is nearly constant. Hence eq. \ref{eq:dL-z} simplifies greatly to, 
\begin{equation} \label{lum_dis_eqn}
   d_L=\frac{cz_{\rm true}}{H_0},
\end{equation}
thus becoming independent of the cosmological model. Incorporating the peculiar velocity (eq. \ref{vp contamination eqn}) into the luminosity distance (eq. \ref{lum_dis_eqn}), we get, 
\begin{equation}
    d_L=\frac{cz_{\rm obs} - v_p}{H_0}.
\end{equation}
Thus, for a bright siren event, with the $d_L$ estimated from the GW data and $z_{\text{obs}}$ estimated from the EM counterpart, the peculiar velocity $v_p$ can lead to a bias in the inferred value of $H_0$ if not accounted for and corrected appropriately.

The peculiar velocity of the galaxies depends on the environment where they are forming. Galaxies that are forming in the dense environment and in large galaxy halo will exhibit larger peculiar velocity (with both linear and non-linear components) \citep{Fry_1994,Song_2009,Lavaux_2011,Hudson_2012,Zheng_2013,Cuesta-Lazaro_2020}. The formation of galaxies in massive halos in a dense environment will have different astrophysical properties such as stellar mass, and star formation rate (see a review by \citealp{2018ARA&A..56..435W}) and as a result, there will exist a correlation between the galaxy properties such as stellar mass and peculiar velocity of the galaxies. 

Along with the connection between the galaxies and halos, the formation of the GW sources is related to the properties of galaxies and hence also properties of the halo. Formation of bright standard sirens and their merging depends on the formation channel and properties of the galaxies such as its stellar mass, star formation rate \citep{Perna:2021rzq, BNS_masses, Santoliquido:2023wzn}. So, depending on the underlying population of the host of GW sources, the halo property of the galaxies and hence the peculiar velocity contamination will be different. In this work, we are interested to show the dependence of the peculiar velocity on the astrophysical property of the host galaxies such as its stellar mass $M_{\star}$ and how the interplay between the underlying GW source population and stellar mass can lead to different amount of contamination from peculiar velocity to the host of the GW sources. A GW source population dependent peculiar velocity contamination will lead to a population dependent impact on the inference of the value of Hubble constant from the low redshift sources. We show the impact of stellar-mass dependent peculiar velocity contamination on the value of Hubble constant for the ongoing and upcoming GW surveys and how to mitigate such effects from future observations with the network of GW detectors such as LIGO+Virgo+KAGRA (\texttt{LVK}) \citep{Harry:2010,Aso:2013,KAGRA:2013rdx,Acernese:2014,LIGOScientific:2014pky, Abbott:2016xvh, KAGRA:2020tym}, LVK+LIGO-India (\texttt{LVKI}) \citep{UNNIKRISHNAN_2013,Saleem:2021iwi}, and Cosmic Explorer+Einstein Telescope (\texttt{CE+ET}) \citep{Punturo_2010, Maggiore:2019uih, Hall_2019, Evans:2021gyd, Adhikari:2022sve}.

\begin{figure*}
    \centering
    \includegraphics[width=\textwidth]{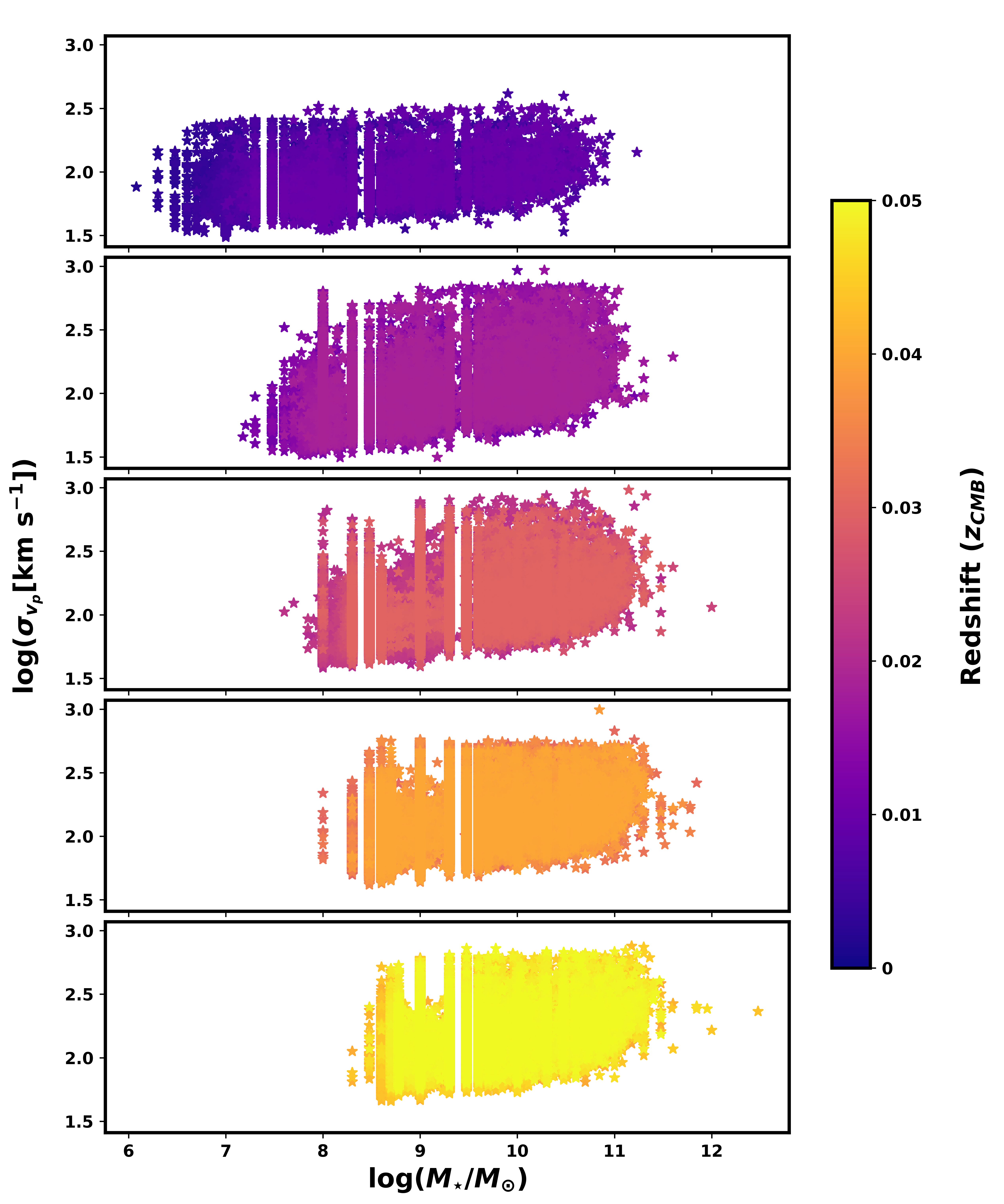}
    \caption{This plot is the entire filtered GLADE+ catalog divided into five subplots based on five redshift bins showing the velocity dispersion as a function of stellar mass.}
    \label{fig:money_plot}
\end{figure*}

\section{Modeling the impact of peculiar velocity} \label{section:vp model}
Efforts for estimating the cosmic peculiar velocity field has spanned decades of extensive research \citep{Kaiser_1991,Shaya_1992,Hudson_1994,Davis_1996,Branchini_1999,Branchini_2001,Nusser_2001, Sheth_2001, Hudson_2004, Radburn-Smith_2004, Pike_Hudson_2005, Lavaux_2011, Davis_2011, Ma_2012, Turnbull_2012, Carrick_2015, Mukherjee_2021}.
In the linear regime, where the density fluctuations are very small, the resulting velocity dispersions are underestimated \citep{Mukherjee_2021}. The peculiar velocity model by \cite{Sheth_2001} accounts for the non-linearities in the density field.

Assuming that all dark matter particles lie in a spherical virialized halo, \cite{Sheth_2001} proposed that the peculiar velocity, $\vec{v}_p$, has two components, 
\begin{equation}
    \Vec{v}_p = \Vec{v}_{halo} + \Vec{v}_{vir},
\end{equation}
where $\Vec{v}_{vir}$ is the higher order term which represents the virial motion of a dark matter particle around the center of mass of parent halo, and $\Vec{v}_{halo}$ represents the motion of the center of mass of the halo. The linear component arises from the bulk flow of a group or cluster while the non-linear term results from the non-linear interactions between galaxies within a cluster. Hence the velocity dispersion for the non-linear component depends upon the mass of the parent halo, $\sigma_{vir} \propto m^{1/3}$. The proportionality constant, obtained from the relations given by \citet{1998Bryan&Norman}, sets the dispersion for the non-linear term as
\begin{equation} \label{sigma_virial}
    \sigma_{vir} = 476g_{\sigma}\big(\Delta_{nl}E(z)^2\big)^\frac{1}{6} \Big(\frac{m}{10^{15}M_{\odot}/h}\Big)^\frac{1}{3} \indent \text{km s}^{-1},
\end{equation}
for a galaxy with a halo of mass $m$ in a region with non-linear overdensity contrast $\Delta_{nl}$ approximated by,
\begin{equation}
    \Delta_{nl} = 18\pi^2 + 60x - 32x^2,
\end{equation}
where $x=\Omega_m(1+z)^3/E(z)^2 - 1$ and the term $g_{\sigma} = 0.9$ is a normalization factor \citep{Sheth_2001}. The fitting form for the bulk flow is obtained by extrapolating root-mean-square velocities of the dark matter particles from peaks in the velocity power spectrum from \cite{Colberg}.

Our work relies on the peculiar velocity estimates from the technique developed by \cite{Mukherjee_2021} which utilizes the \texttt{BORG} (Bayesian Origin Reconstruction from Galaxies) formalism \citep{jasche2013borg, BORG-PM}. By fitting a dynamical structure formation model to observed galaxies in cosmological surveys, the \texttt{BORG} method infers a physically feasible and probabilistic model of the three-dimensional cosmic matter distribution, providing the linear and partially non-linear components of the velocity field. The \texttt{BORG} algorithm accounts for unknown galaxy bias and incorporates selection and evolutionary effects while providing the velocity field as part of the dynamical model. \texttt{BORG} method provides a numerical approximation of the posterior distribution of the parameters in a spatial grid of $256^3$ values with a spatial resolution of 2.64 Mpc $h^{-1}$ for the initial conditions plus the bias parameters. The dark matter particle's initial and final positions are provided for each sample of the posterior, allowing for the estimation of the velocity field using the Simplex-in-Cell estimator.

The implementation of \texttt{BORG} framework has been discussed extensively in \cite{Mukherjee_2021}. \texttt{BORG} reconstructs the initial density field and simulates the evolution of the density fluctuations under gravity via linear perturbation theory to estimate the $v_{halo}$ (linear) component. This component paired with the non-linear term (eq. \ref{sigma_virial}) together forms the peculiar velocity field.

In order to explicate the correlation between the peculiar velocity of host galaxies of GW sources and their properties, we consider real galaxies from the GLADE+ galaxy catalog \citep{GLADEplus} (more details in section \ref{subsection:GLADE+}). GLADE+ catalog consists of the redshifts of the galaxies corrected for their peculiar motions using the \texttt{BORG} formalism discussed above. By filtering the catalog to include galaxies with $z_{\text{true}}\leq0.05$, we examine the distribution of the peculiar velocity as a function of stellar mass. 

Fig. \ref{fig:money_plot} shows the distribution of the log of peculiar velocity dispersions as a function of the log-stellar mass of all galaxies up to redshift $z=0.05$ from the GLADE+ galaxy catalog. The total distribution is divided into five sub-distributions based on five redshift ranges ranging from $z= 0.01$ to $z=0.05$. As indicated by the color bar, the topmost panel shows the distribution for $0<z\leq0.01$, the second panel shows the same for $0.01<z\leq0.02$, and so on. As we proceed from the top panel towards the bottom, more galaxies are detected with higher stellar mass. This is because the galaxies that are further away with less stellar mass are fainter and were not detected in a magnitude-limited survey. The number of galaxies detected is also large as the comoving volume of the Universe at high redshift is large. Regardless of the redshift range, an increase in the stellar mass leads to an increase in the peculiar velocity dispersion.

\begin{figure*}
    \centering
    \includegraphics[width=\textwidth]{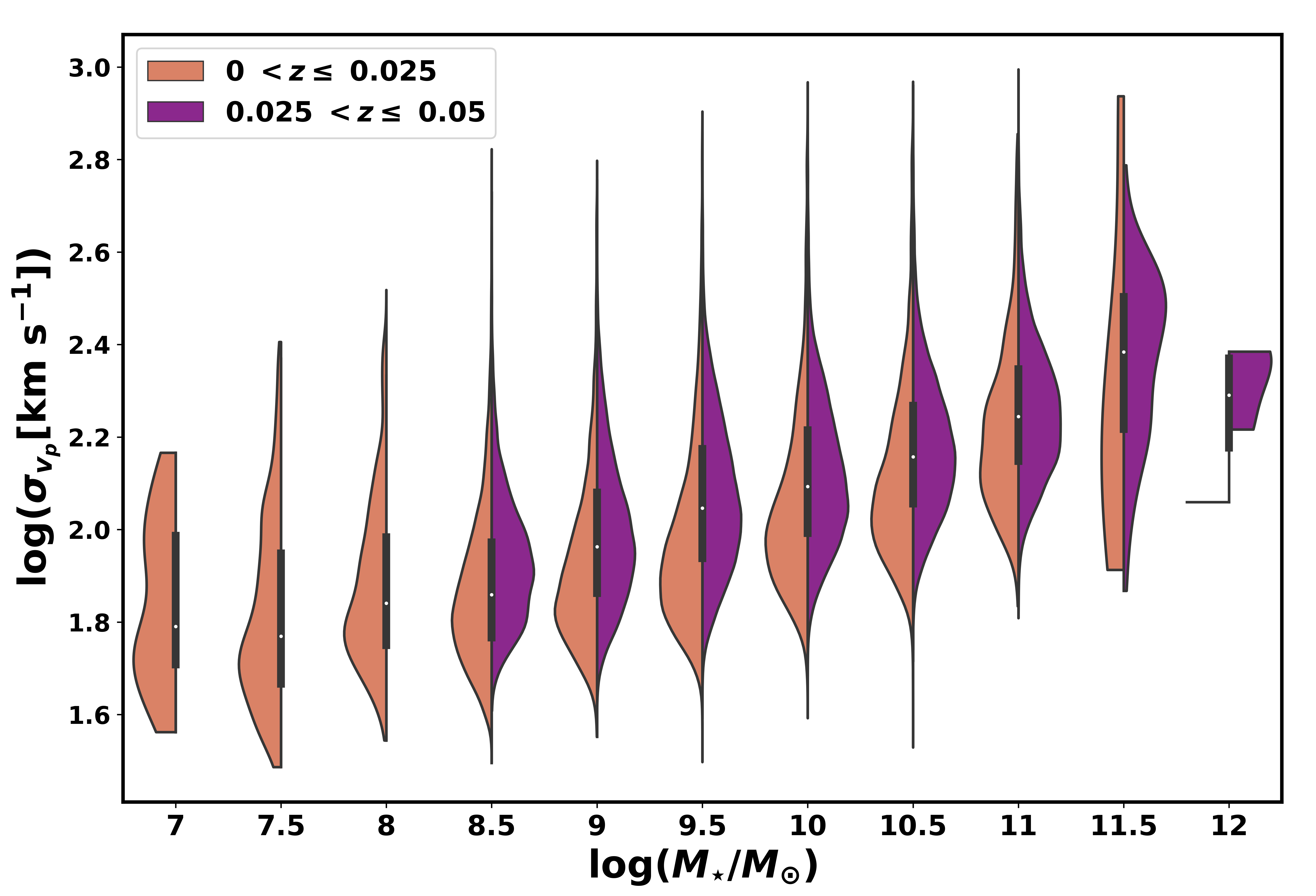}
    \caption{The median distribution of the log-peculiar velocity as a function of log-stellar mass for all galaxies up to redshift of 0.05 from the GLADE+ catalog. The two posteriors in each violin represent the two redshift bins.}
    \label{fig:violin_plot}
\end{figure*}

To show the dependence of the dispersion of the peculiar velocity on the host property of the galaxy, we show a violin plot in Fig.  \ref{fig:violin_plot}, which represents the median distribution of the log of the peculiar velocity dispersion as a function of log-stellar mass of galaxies. Here the galaxies in each stellar mass bin are divided into two redshift bins, $z$ between 0 and 0.025 and $z$ from 0.025 up to 0.05. For each stellar mass bin, the orange curves represent the distributions for galaxies in the lower $z$ bin ($0<z\leq 0.025$), and the purple ones represent those in the higher $z$ bin ($0.025<z\leq 0.05$). The first three violins do not exhibit a purple posterior which indicates that for those stellar mass bins, there are no galaxies detected in the higher $z$ bin as low-mass galaxies are not bright enough to be detected at higher redshifts. This can be validated from the lower panels of Fig.  \ref{fig:money_plot}. Overall, we see an increasing trend in the median distribution of log-peculiar velocity dispersion with an increase in the order of the stellar mass in both the redshift bins, which indicates a population-dependent peculiar velocity contamination. {In every stellar mass bin, the orange posteriors are skewed towards a lower value of peculiar velocity as compared to the purple posteriors which indicates a redshift-dependent contamination from peculiar velocity.} 
Previous studies on other datasets have also found a similar trend of increase in the velocity dispersion with an increase in the stellar mass \citep{2017ApJS..229...20S, 2020MNRAS.498.5704N}. The velocity trend observed from GLADE+ agrees with previous findings. The key summary from this section is that the underlying peculiar velocity dispersion depends on the stellar mass and host redshift of the sources. As a result, whether most GW sources are hosted in galaxies with higher (or lower) stellar mass, their peculiar velocity contamination will be different. It is important to note here that this impact can be also important for standard candles. These sources can also exhibit a population-dependent peculiar velocity contamination. We will explore this in a future work. 

\section{Modeling GW source population for bright sirens} \label{section:GW source population modelling}
Bright standard sirens have an observable EM counterpart and hence are expected to be detected at low redshifts. The contribution of peculiar velocity is significant typically up to $z=0.05$ \citep{Mukherjee_2020measurement}. Our work focuses on low redshift bright sirens, particularly BNS mergers. To probe the impact of the host properties of BNSs, it is essential to model their population appropriately. The merger rate of compact objects plays a key role in this aspect. The merger rate for compact binaries as a function of some physical property like stellar mass or redshift gives the probability distribution for their host galaxies with respect to that property.

\subsection{Merger rate density distribution for BNS hosts} \label{subsection:stellar_mass_distribution}
According to our current understanding of galaxy formation, the formation and growth of a galaxy depends on the halo properties [a review article on this topic \citep{Wechsler:2018pic}]. The relation between the halo-mass and stellar mass in a galaxy as a function of redshift is a useful indicator for the formation of galaxies. Almost all properties of galaxies are strongly influenced by their stellar mass \citep{Kauffmann_galaxy_properties,Taylor_2011}. High (low) mass galaxies typically exhibit old (young) stellar populations, high (low) mass-to-light ratios, and low (high) star formation rates \citep{Kauffmann_2003b}. There is also a tight correlation between stellar mass and the metal content in the gas phase of emission-line galaxies \citep{Kauffmann_2003a,Tremonti_2004} indicating that the amount of metals in a galaxy is closely connected to its stellar mass.

The formation of BNSs in a galaxy depends on the halo mass and also the stellar properties of a galaxy such as stellar mass, star formation rate, stellar metallicity \citep{Perna:2021rzq, BNS_masses, Santoliquido:2023wzn}. As a result, the merger rate of BNSs at a redshift depends on the host galaxy properties such as the stellar mass, which in turn also depends on the halo-mass. The properties of galaxies in which binary compact objects (BCOs) form (called formation galaxies) can significantly vary from those in which BCOs merge (called host galaxies). The properties of the host galaxy and formation galaxy can differ significantly depending on the growth history of the galaxy and also the delay time between the formation of the stars and the merging of a BNS. For instance, the formation and evolution of a BNS system span $10^6$ years (1 Myr) to $10^9$ years (1 Gyr) before it finally merges. Throughout this timescale, its formation galaxy can be subjected to significant alterations via chemical evolution or even galaxy-galaxy merger resulting in its host galaxy having properties dissimilar to its formation galaxy.

To model the host properties of GW sources, \citet{BNS_masses} simulated the mergers of compact binaries using binary population models for their progenitors while considering the evolution and possible mergers of their host galaxies across different cosmic timescales. The host properties such as the galaxy stellar mass, metallicity and star formation rate (SFR) are assumed to be following either the mass metallicity relation (MZR) \citep{Tremonti_2004, Kewley_2008, Maiolino_2008, Mannucci_2009, Magnelli_2012, Zahid_2014, Genzel_2015, Sanders_2019} or the fundamental metallicity relation (FMR) \citep{Mannucci_2010, Mannucci_2011, Hunt_2012, Hunt_2016, Curti_2019, Sanders_2019}. As seen in section \ref{section:vp model}, the peculiar velocity dispersion depends upon the mass of the parent halo of the host galaxy and hence on their stellar mass content (see Fig. \ref{fig:violin_plot}). To model the source population, we adapt from \citeauthor{BNS_masses} the distribution of BNS merger rate density as a function of the host stellar mass for local galaxies assuming FMR. Following their log-normal merger rate density distribution (\citeauthor{BNS_masses} Page 9, figure 8: BNS $\alpha1$ FMR), we fit a broken power law (more in section \ref{subsection:GLADE+}) using the following equation
\begin{equation} \label{equation:broken_power_law}
    f(dm) \propto \Big(\frac{dm}{m_{\text{break}}}\Big)^{\alpha_1} \Big\{\frac{1}{2} \Big(1 + \Big(\frac{dm}{m_{\text{break}}} \Big)^{\frac{1}{\Delta}} \Big) \Big\}^{(\alpha_2 - \alpha_1) \Delta}
\end{equation}
where $dm = $ dlog(M$_{\star}$/M$_{\odot}$), $m_{\text{break}}$ is the pivotal point of change of slope, $\alpha_1$ and $\alpha_2$ are the power law indices and $\Delta$ is the smoothness parameter. Normalizing the above equation for the probability density function (PDF), we get the probability densities for BNS host galaxies across a log-stellar mass range of [6, 13]  which gives the probability that a GW event detected by a detector is emitted by a coalescing BNS located in a galaxy that has log-stellar mass given by the log-mass bin $dm$. The value of $m_{\text{break}}$ in the work by \cite{BNS_masses} is approximately $10^{10.5}$ M$_\odot$ which implies that galaxies with a stellar mass of the order of 10$^{10}$ M$_{\odot}$ are estimated to have the largest number of BNS mergers per Gpc$^{3}$ yr$^{-1}$ (following the merger rate distribution).

\begin{figure}
    \centering
    \includegraphics[width=0.48\textwidth]{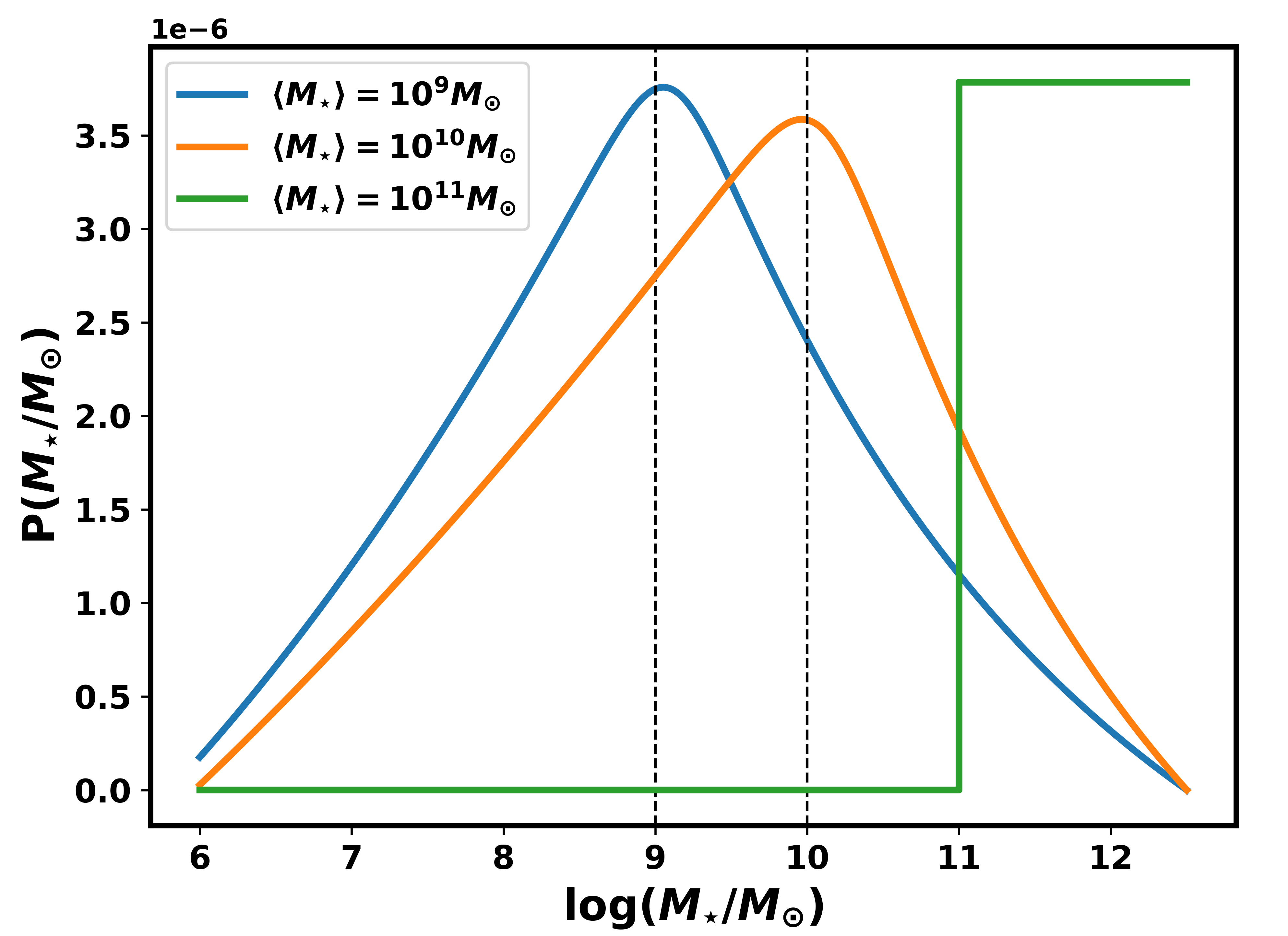}
    \caption{The probability distributions over log-stellar mass used to sample three different populations of galaxies from the filtered GLADE+ catalog. The blue and orange curves are fitted with a broken power law with the given parameters. The green line is a step function used to restrict samples to high-mass galaxies.}
    \label{fig:mass_distribution}
\end{figure}

\subsection{Redshift dependent merger rate model} \label{section:redshift distribution}
\begin{figure}
    \centering    
    \includegraphics[width=0.48\textwidth]{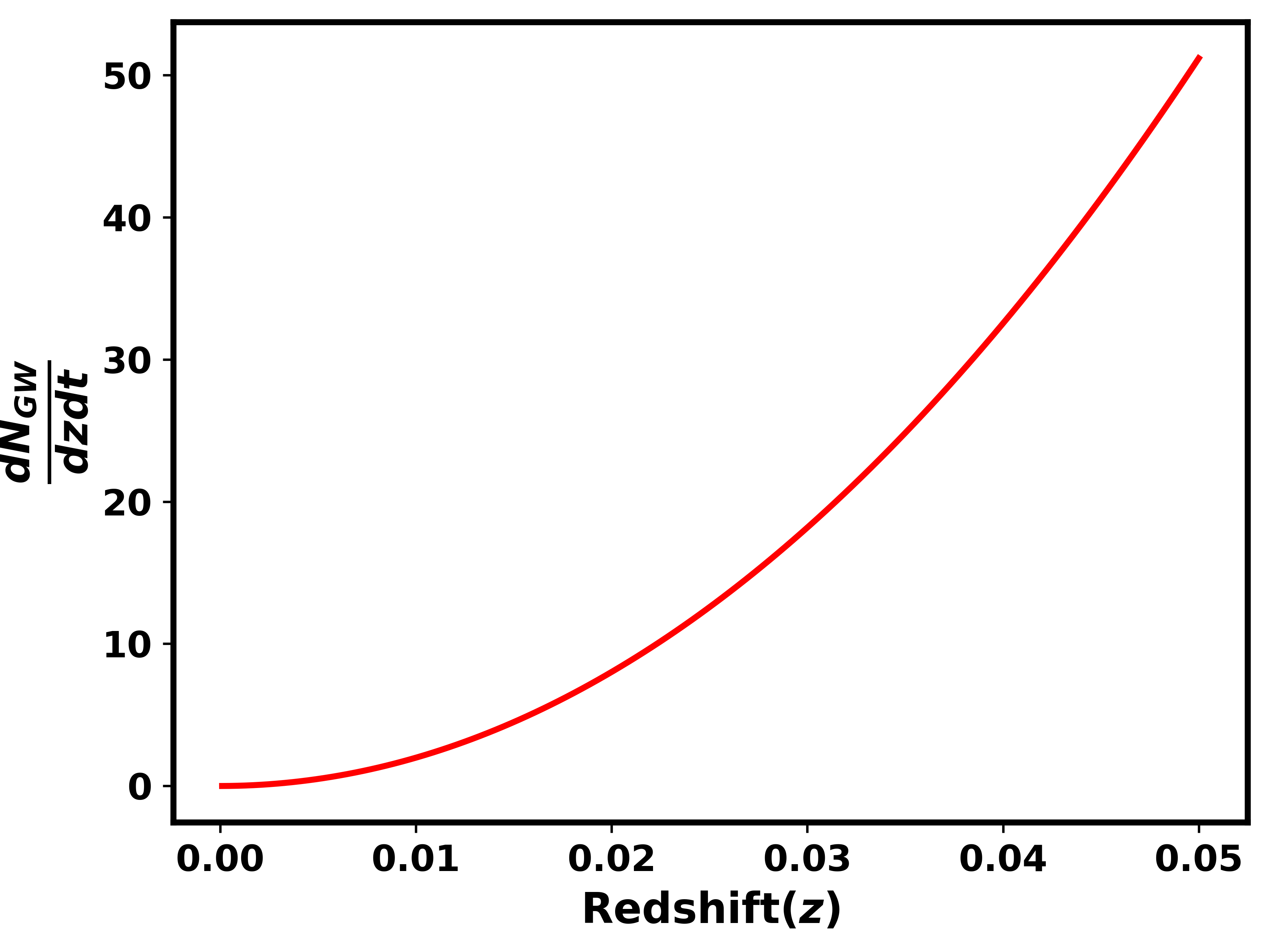}
    \caption{The number of BNS coalescing events per redshift bin $dz$ for a given observation time \textit{dt}. The curve shows the distribution of the host galaxies of GW sources up to a redshift $z = 0.05$.}
    \label{fig:Redshift distribution}
\end{figure}

Having constructed a model describing the distribution of stellar masses across the galaxies hosting BNS systems, the next stage involves modeling the redshifts associated with these galaxies. The number of Compact Binary Coalescing (CBC) events per unit redshift per unit observation time is estimated using \citet{GWSim} as
\begin{equation} \label{dndzdt}
    \frac{dN_{\rm GW}}{dz dt} = \frac{R(z)}{1+z} \frac{dV_c}{dz} \big(\theta_{\Lambda \text{CDM}}\big);
\end{equation}
where $\frac{dV_c}{dz}\big(\theta_{\Lambda \text{CDM}}\big)$ is the  comoving volume at redshift $z$ for the cosmological parameters denoted by $\theta_{\Lambda \text{CDM}}$ for the flat-$\Lambda$CDM model and $R(z)$ is the redshift evolution of the merger rate for different delay time distributions 
\begin{equation}
    R(z) = R_0 \int_z^\infty P(t_d|t_d^{min},t_d^{max},d) R_{SFR}(z_m)\frac{dt}{dz_m}dz_m,
\end{equation}
where {$P(t_d|t_d^{min},t_d^{max},d) = (t_d)^{-d}$ is the probability distribution of the delay time ($t_d$) given the power law index $d$; $R_{SFR}(z)$ is the star-formation rate at redshift $z$ \citep{Madau_Dickinson} and R$_0$ is the merger rate at redshift $z=0$. The minimum and maximum time delay is given in terms of the lookback time and the local merger rate R$_0$ for BNS is inferred by \cite{ligo_2022_population} between 10 Gpc$^{-3}$ yr$^{-1}$ and 1700 Gpc$^{-3}$ yr$^{-1}$. Our work assumes a fiducial local merger rate of R$_0 = 20$ Gpc$^{-3}$ yr$^{-1}$. Hence eq. (\ref{dndzdt}) gives the number of BNS coalescing events per unit redshift per unit observation time. This gives the distribution of BNS host galaxies per redshift bin $dz$ for a given observation time $dt$.
Integrating the eq. (\ref{dndzdt}) for a given observation time (including the detector duty cycle) and the redshift range gives the total number of BNS events up to that redshift. Hence, for redshifts up to 0.05, the distribution of BNS host galaxies is shown in Fig.  \ref{fig:Redshift distribution}.} In this analysis we consider the BNS to follow the Madau-Dickinson SFR \citep{Madau_Dickinson} with a negligible minimum delay time. 

\subsection{GLADE+ galaxy catalog} \label{subsection:GLADE+}  
\begin{figure}
    \centering
    \includegraphics[width=0.48\textwidth]{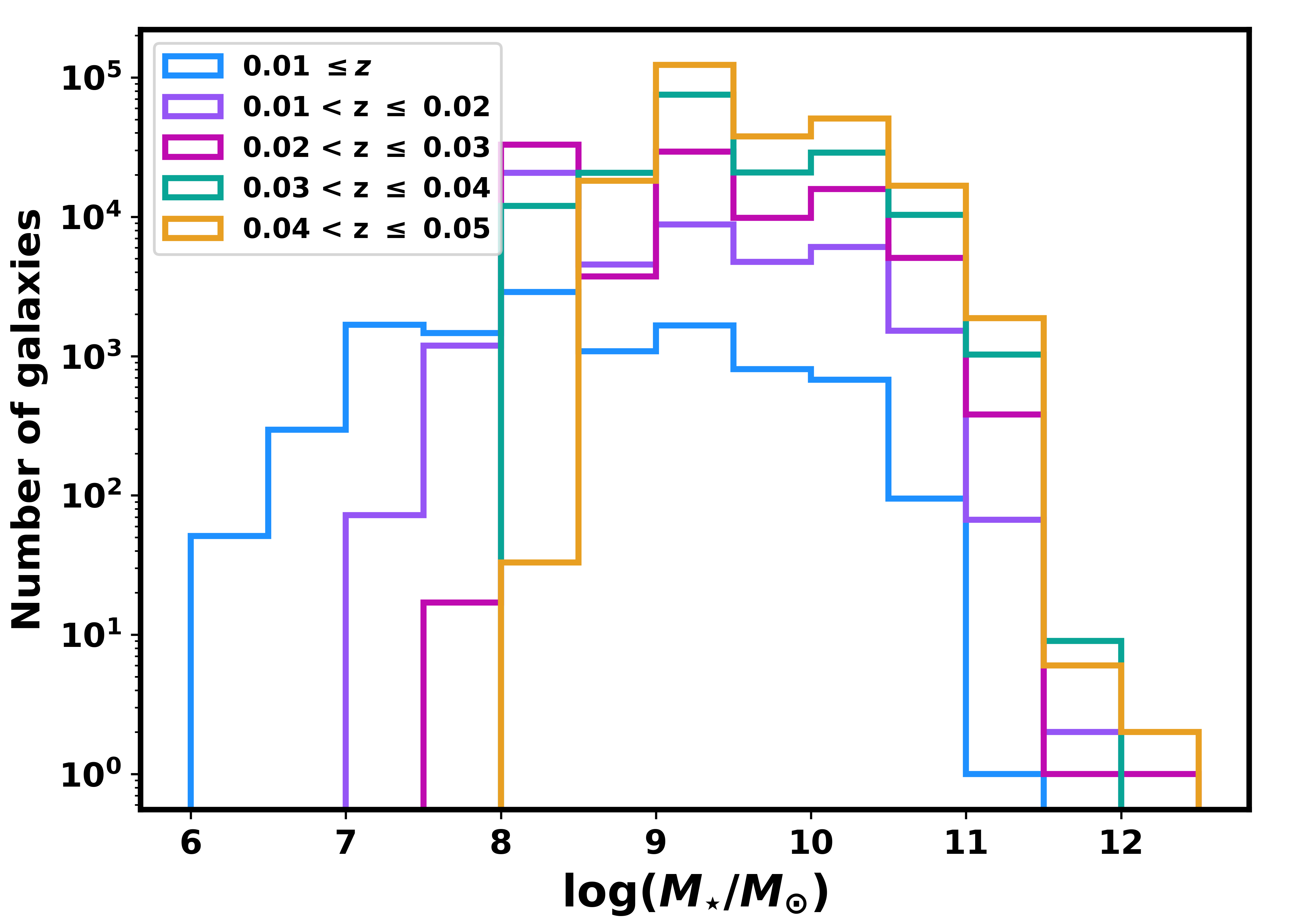}
    \caption{Distribution of galaxies from the filtered GLADE+ catalog as a function of their log-stellar mass. The distribution includes galaxies up to $z=0.05$ in 5 bins.}
    \label{subfig:glade+mass}
\end{figure}

Using the mass distribution model of BNS and merger rate model, we choose the host galaxies of BNS from the GLADE+ galaxy catalog \citep{GLADEplus}. In order to do so, we consider the distribution of host galaxies of BNSs across the stellar mass range of $10^6$ M$_{\odot}$ to $10^{12}$ M$_{\odot}$ along with their distribution across redshift up to $z = 0.05$. The GLADE+ galaxy catalog is constructed with the purpose of optimizing multi-messenger searches by facilitating the process of sky localization for EM follow-up. The catalog contains 22.5 million galaxies and 7,50,000 quasars. It consists of the redshifts of the galaxies in the CMB frame, their stellar masses, and their peculiar velocity uncertainties estimated from the BORG framework \citep{Mukherjee_2021, jasche2013borg, BORG-PM}. The redshift flag (z\_flag) with a value of 1 indicates that the redshift of the galaxy in the CMB frame is corrected for the peculiar velocity bias, i.e., $z_{\text{CMB}}$ = $z_{\text{true}}$.
Hence, to garner a population of potential BNS host galaxies, we filter the GLADE+ catalog by applying the following conditions:
\begin{itemize}
    \item $z_{\text{CMB}} \le 0.05$,
    \item M$_{\star} \neq$ `null',
    \item z\_flag = 1.
\end{itemize}
With these restrictions, we get a subset of GLADE+ (henceforth called `filtered GLADE+') containing 572558 galaxies. The filtered GLADE+ contains galaxies up to $z \leq 0.05$ corrected for their peculiar velocity bias and transformed to the CMB frame. It also contains the stellar masses ($M_{\star}$), peculiar velocity dispersion ($\sigma_{v_p}$), and the sky location (right accession (RA) and Declination (Dec)) of galaxies up to $z_{\text{CMB}} = 0.05$. Fig. \ref{subfig:glade+mass} shows the distribution of galaxies from the filtered GLADE+ catalog as a function of the stellar mass.

To understand the impact of the host galaxy properties of BNSs on the peculiar velocity and subsequently, on the inference of $H_0$, we consider three different cases of BNS population models to sample potential BNS hosts. The host galaxies from filtered GLADE+ are chosen using these three different population models: 
\begin{enumerate}
    \item $\langle M_{\star}\rangle = 10^9 M_{\odot}$,
    \item $\langle M_{\star}\rangle = 10^{10} M_{\odot}$,
    \item $\langle M_{\star}\rangle = 10^{11} M_{\odot}$.
\end{enumerate}

We have considered these mass models by adopting the merger rate density distribution curve from \cite{BNS_masses} and shifting the pivotal point of the stellar mass. As explained in section \ref{subsection:stellar_mass_distribution}, these mass models are log-normal distributions and are fitted using broken power law given by eq. \ref{equation:broken_power_law} with the parameters given in table \ref{tab:mass_broken_powerlaw_params}. The broken power law forms a PDF which then is combined with the normalized densities from Fig. \ref{subfig:glade+mass} to sample potential BNS host galaxies from filtered GLADE+. The blue and orange curves in Fig. \ref{fig:mass_distribution} are the PDFs corresponding to the parameters given in table \ref{tab:mass_broken_powerlaw_params}. The third case (green curve) however, is not fitted with a broken power law because if the pivotal point is set at $10^{11} M_{\odot}$, the normalized density from Fig. \ref{subfig:glade+mass} dominates the broken power law as there are less than 1\% galaxies having $M_{\star} \geq 10^{11} M_{\odot}$. Hence, for $\langle M_{\star}\rangle = 10^{11} M_{\odot}$ case, we use a step function to restrict the population to masses $M_{\star} \geq 10^{11} M_{\odot}$.

\begin{table}
    \centering
    \begin{tabular}{|c|c|c|}
         Parameter&$\langle M_{\star}\rangle = 10^9 M_{\odot}$&$\langle M_{\star}\rangle = 10^{10} M_{\odot}$ \\\cline{1-3}
         $m_{\text{break}}$&9&10\\
         $\alpha_1$&2.4&1.7\\
         $\alpha_2$&-3.4&-4.6\\
         $\Delta$&0.015&0.010\\\cline{1-3}
    \end{tabular}
     \caption{Parameters for the broken power law distributions on stellar mass for the two of the three mass models used for sampling sources from the filtered GLADE+ catalog. The parameters for the third case are not included as it is fitted with a step function instead of a broken power law.}
    \label{tab:mass_broken_powerlaw_params}
\end{table}

\begin{figure}
    \centering
    \includegraphics[width=0.48\textwidth]{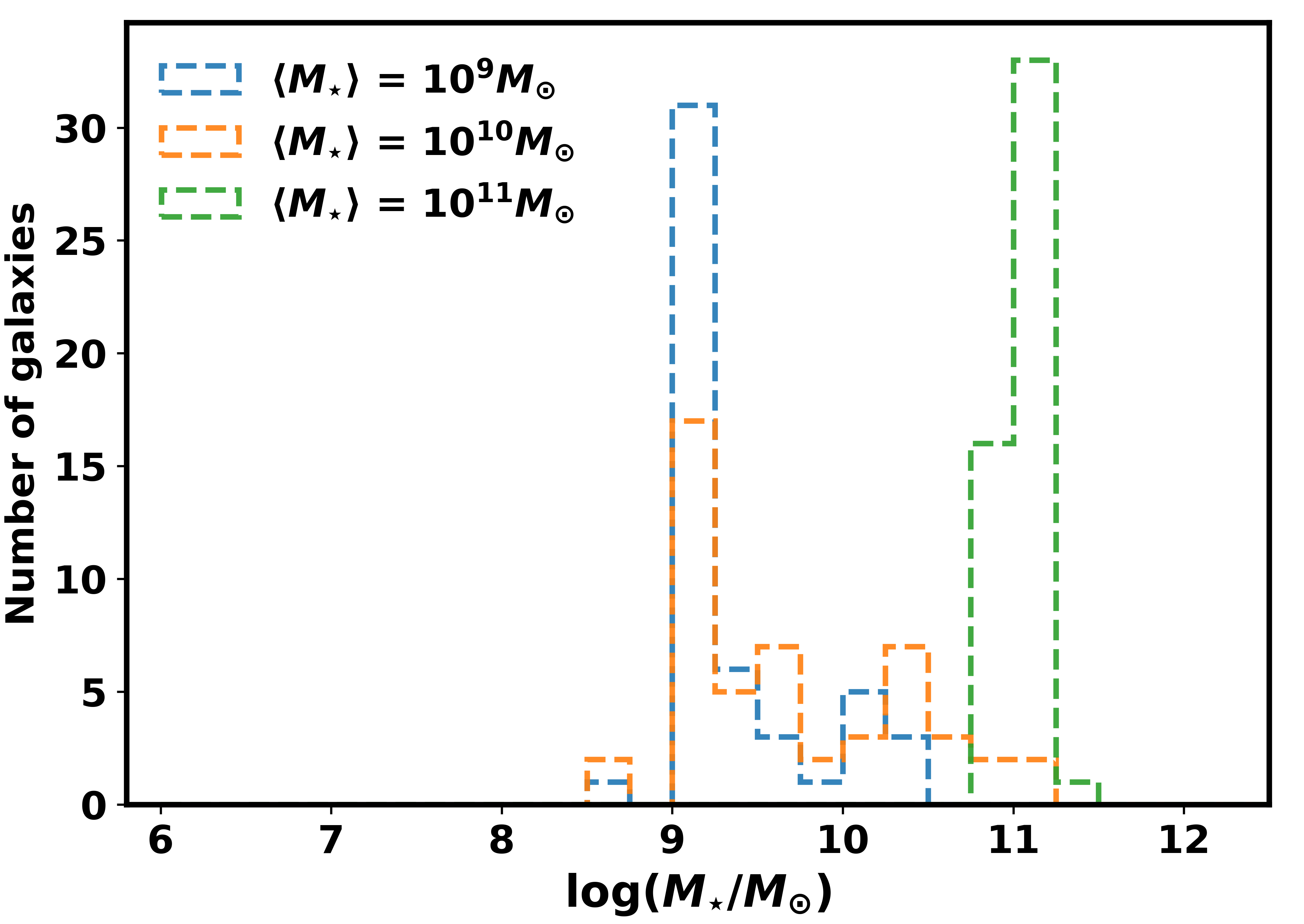}
    \caption{Distribution of the three populations of 50 sources chosen from the GLADE+ catalog with the population model discussed in section \ref{section:GW source population modelling}. This distribution represents galaxies chosen for parameter estimation in section \ref{subsection:BNS source PE}.}
    \label{fig:selsources}
\end{figure}

Using this combined probability distribution on the stellar masses, we sample a population of 50 potential BNS host galaxies per mass model from the filtered GLADE+ catalog that follow the BNS merger rate model from \cite{GWSim} discussed in section \ref{section:redshift distribution}. The corresponding distribution of the sampled galaxies for the three mass cases is shown in Fig. \ref{fig:selsources}. Note that the $\langle M_{\star}\rangle = 10^{11} M_{\odot}$ population contains only high-mass galaxies and the use of the step function does not imply that the mean stellar mass of this population is $10^{11} M_{\odot}$. We use the symbol ``$\langle M_{\star}\rangle$'' as a convention to denote the galaxy populations throughout this paper. 

Thus, we have three populations of potential BNS host galaxies for the three $\langle M_{\star} \rangle$ cases. These three populations have their masses as discussed above, host redshifts that follow the merger rate model, and their peculiar velocity dispersion  ($\sigma_{v_p}$) for individual galaxies. With these velocity dispersions and host galaxy redshift corrected for peculiar velocity we get the $v_p$ realizations for each galaxy in every population.

\subsection{Luminosity distance estimation for BNS sources} \label{subsection:BNS source PE}
To estimate the value of the Hubble constant from these sources, we need an estimation of the luminosity distance for these mock BNS samples for all three different detector network configurations chosen in this analysis. To estimate the luminosity distance, we follow two different procedures, (i) we use the parameter estimation code Bilby for 10 BNS sources, and (ii) we estimate the posterior on the luminosity distance using a Gaussian approximation for 50 sources with a standard deviation equal to the median uncertainty on luminosity distance estimate from Bilby luminosity distance estimates. As we are using median errors on distance estimated from Bilby, we take into account the degeneracy with the inclination angle in a Gaussian approximation. However, this approximation cannot capture the non-Gaussian nature of the posterior. However, on combining multiple GW sources (about 10 sources) for estimating $H_0$, the posterior on Hubble constant $H_0$ will become Gaussian by central limit theorem. As a result, this approximation will not impact the conclusion significantly (this is discussed in detail later). We do not perform Bilby parameter estimation for 50 GW sources to reduce the computational cost.

\begin{figure}
    \centering
    \begin{subfigure}
        \centering
        \includegraphics[width=0.48\textwidth]{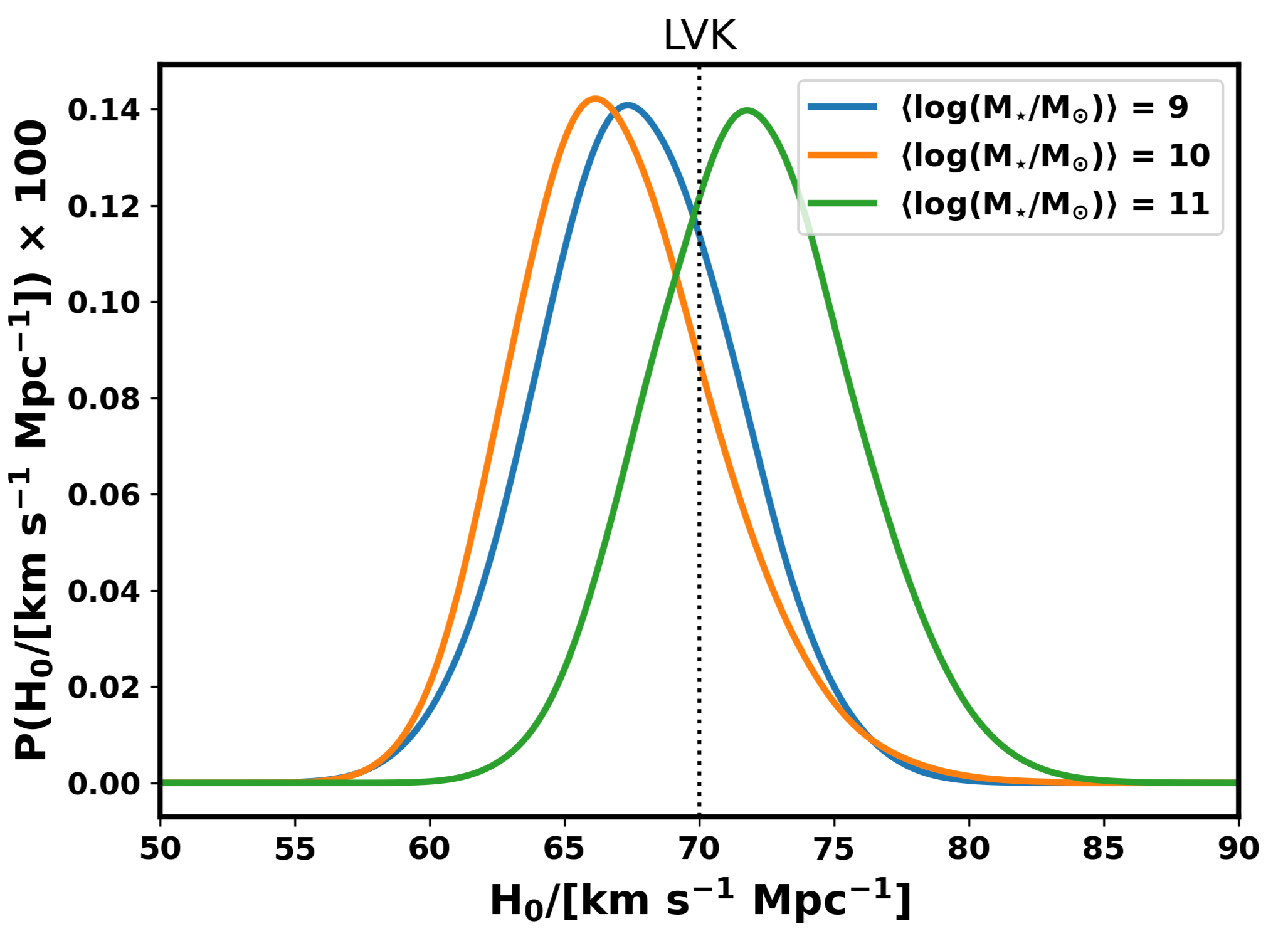}
    \end{subfigure}

    \begin{subfigure}
        \centering
        \includegraphics[width=0.48\textwidth]{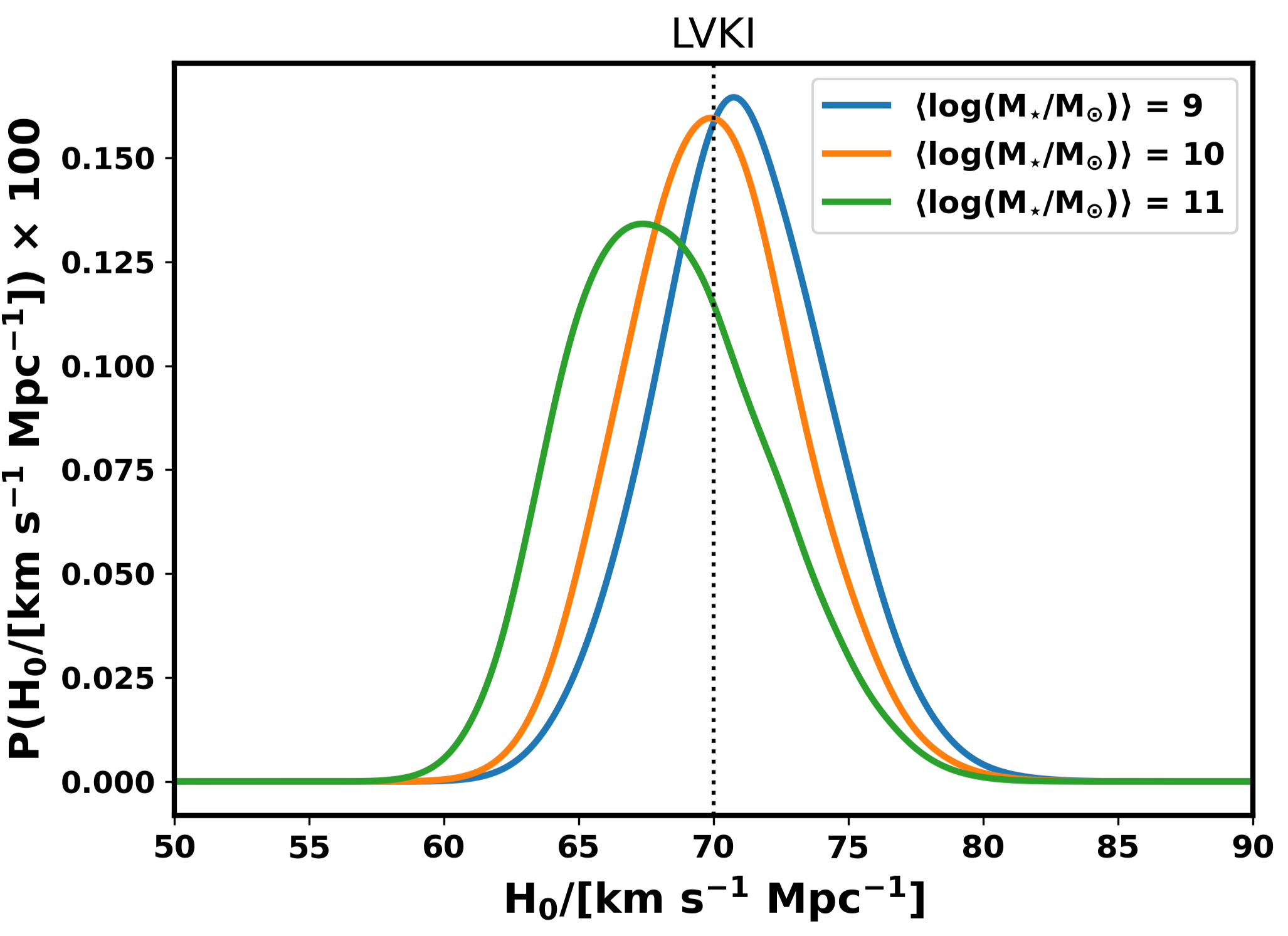}
    \end{subfigure}
    
    \begin{subfigure}
        \centering
        \includegraphics[width=0.48\textwidth]{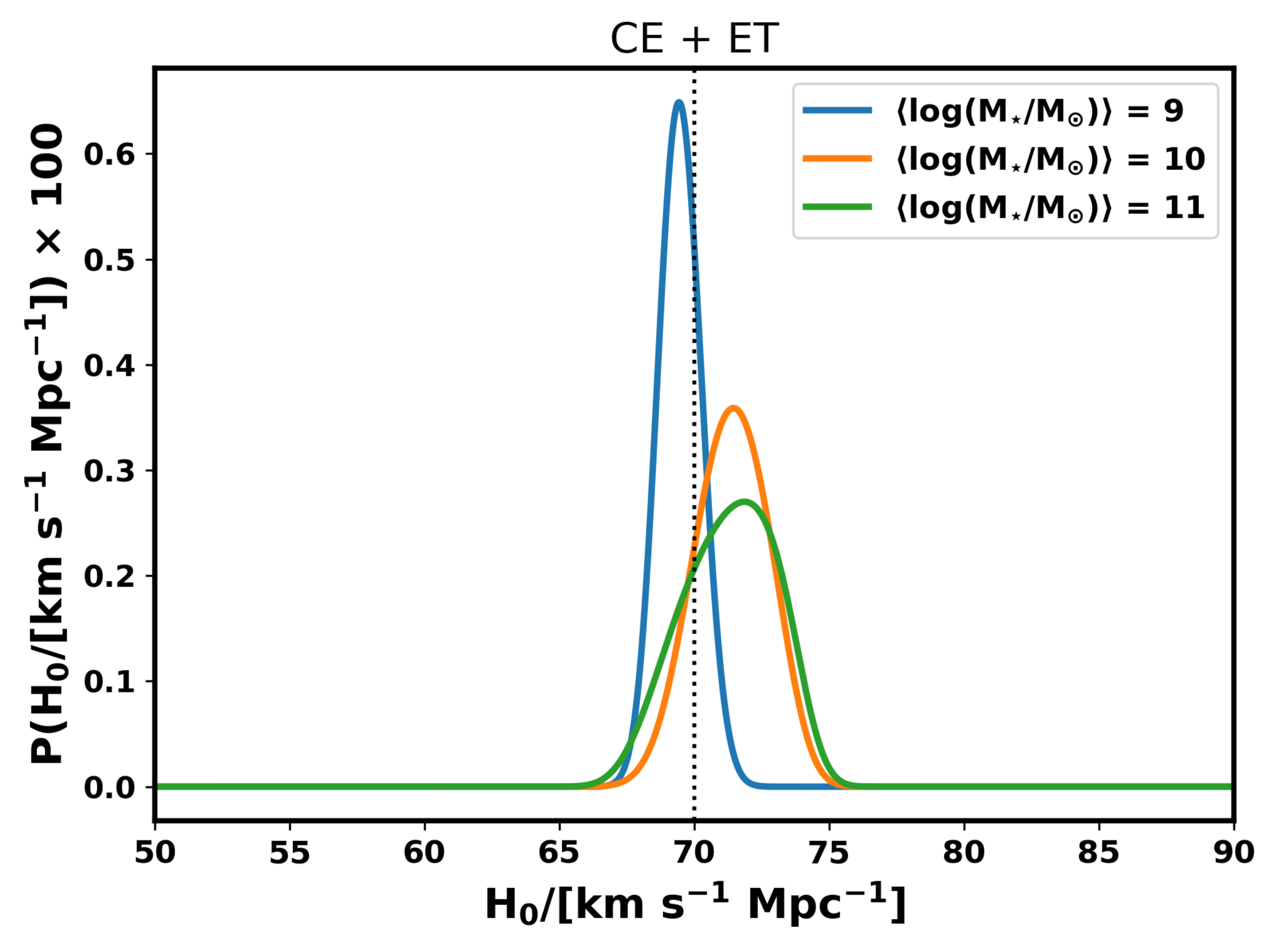}
    \end{subfigure}
    \caption{The combined posteriors of $H_0$ obtained for 10 simulated BNS merger events after correcting for the $v_p$ bias for individual sources from the three BNS host populations from GLADE+. The three panels represent the combined $H_0$ posteriors for the three detector combinations. Each panel contains three posteriors for the three $\langle M_{\star} \rangle$ populations.}
    \label{fig:H0_posteriors}
\end{figure}

\begin{figure}
    \centering
    \includegraphics[width=0.5\textwidth]{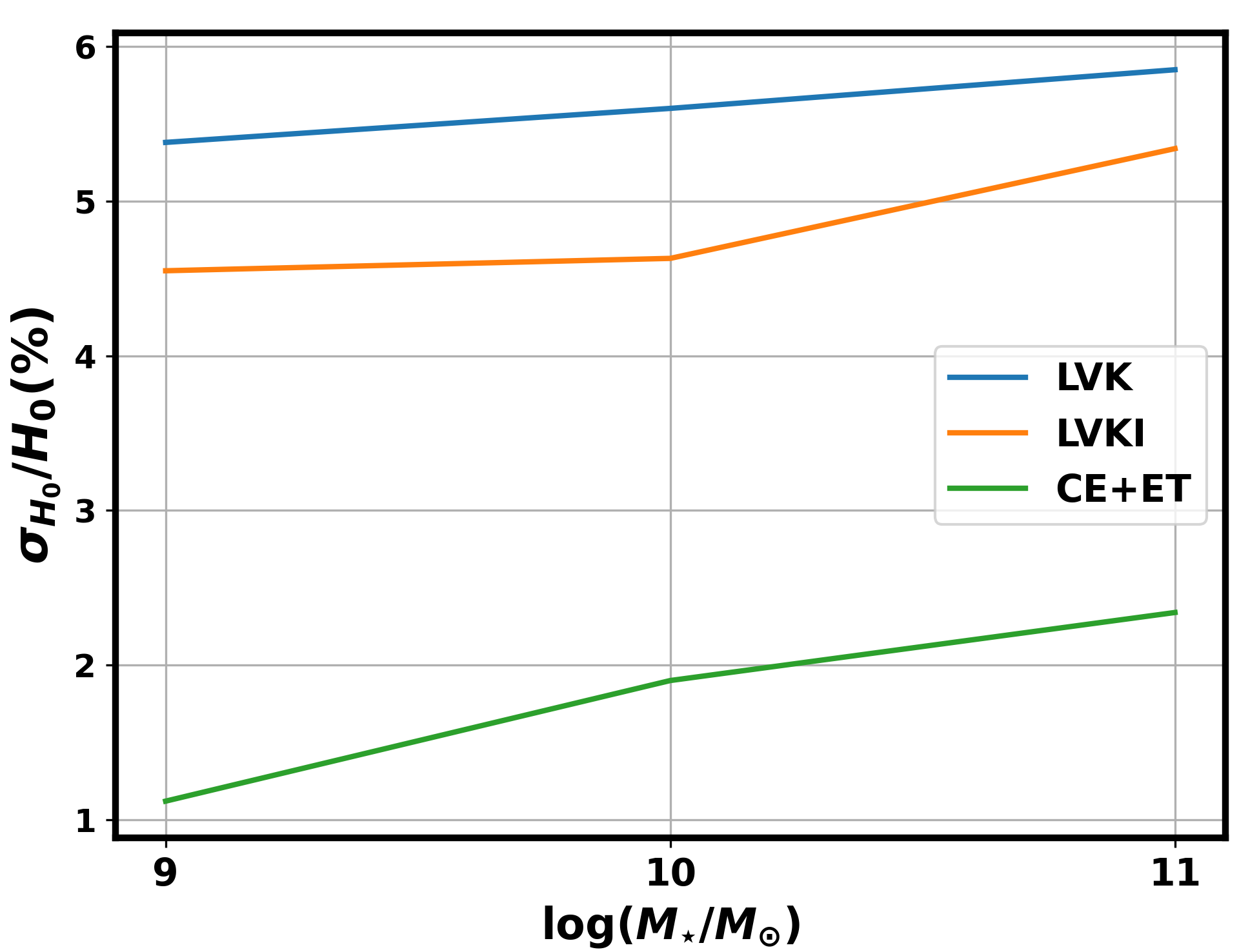}
    \caption{The precision on $H_0$ as a function of the log-stellar mass for the three detector configurations, obtained after combining posteriors from 10 BNS merger events, each simulated in the chosen population of galaxies from the GLADE+ catalog. The corresponding luminosity distance posteriors are obtained from parameter estimation.}
    \label{fig:error_stellar_mass_bilby}
\end{figure}

\textit{Parameter estimation using Bilby: } We perform parameter estimation of the GW sources using the \texttt{Bilby} \citep{Ashton_2019} package to get a realistic posterior on the luminosity distance for the BNSs for three different detector networks, namely:
\begin{enumerate}
    \item \texttt{LVK}: LIGO-Livingston + LIGO-Hanford + Virgo + KAGRA
    \item \texttt{LVKI}: LVK + LIGO-India
    \item \texttt{CE + ET}: Cosmic Explorer + Einstein Telescope
\end{enumerate}
To begin with, we first simulate BNS merger events in the chosen galaxies from the three galaxy populations from the GLADE+ catalog by injecting the GW source parameters $m_1, m_2, d_L, \theta_{\text{JN}}, \psi, \phi$, RA, and Dec. The parameters $m_1$ and $m_2$ denote the component masses randomly sampled from a uniform distribution over the range [0.8 M$_{\odot}$, 2.5 M$_{\odot}$]. The injected $d_L$ is calculated using eq. \ref{lum_dis_eqn} where $z_{\text{true}} \equiv z_{\text{CMB}}$ from GLADE+ and the injected value of $H_0$ is 70 km s$^{-1}$ Mpc$^{-1}$. The sky location of the injected GW source, given by the right ascension (RA) and declination (Dec), is obtained from the GLADE+ catalog. The injected values for the inclination angle $\theta_{\text{JN}}$ and the polarisation angle $\psi$ are randomly sampled from uniform distributions over [0, $\pi$] while that for the GW phase $\phi$ is sampled from uniform over [0, 2$\pi$]. We do not consider any spin effects like precession or tidal deformation in this analysis.

\begin{figure*}
    \centering
    \includegraphics[width=\textwidth]{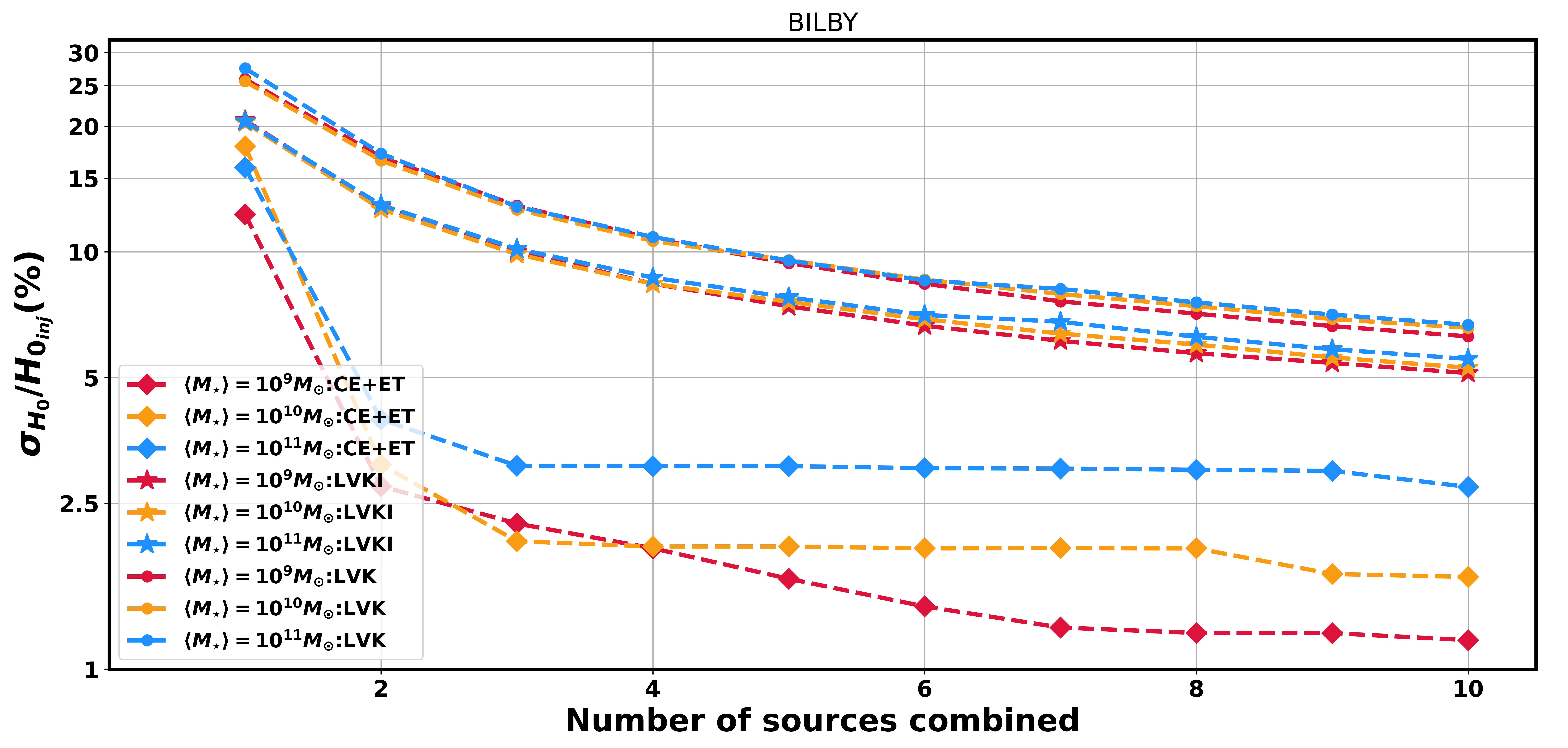}
    \caption{Precision on $H_0$ as a function of the number of sources cumulatively combined for the three detector configurations (denoted by distinct markers) for the three $\langle M_{\star} \rangle$ populations (distinctly colored). The distance posteriors are obtained from parameter estimation discussed in section \ref{subsection:BNS source PE}.}
    \label{fig:combined glade_error 10 to 50}
\end{figure*}

With these injection parameters, a GW signal waveform is generated using the \texttt{IMRPhenomPv2} waveform model. With a sampling rate of 2GHz, the frequency domain waveform $h(f)$ is then added to the Gaussian colored noise $n(f)$ with the power spectral density $S_n(f)$ for the design sensitivity of LIGO, Virgo, KAGRA, LIGO-India, Cosmic Explorer, and Einstein Telescope. This generates the GW strain data for the injected source parameters. The data duration for each source for the \texttt{LVK} and \texttt{LVKI} cases is $\sim 15 - 20$ minutes varying with the randomly sampled component masses. The data duration for \texttt{CE+ET} case is $\sim 25 - 35$ minutes.

Since the luminosity distance $d_L$ and the inclination angle $\theta_{JN}$ of the source are the only parameters of interest, we set the priors for all the other parameters as delta functions to expedite the parameter estimation. This approximation is appropriate as the luminosity distance and masses are not degenerate with these detector configurations. Luminosity distance and inclination angle are the maximally degenerate parameters, so we primarily consider these two for the parameter estimation. For $d_L$, the prior is uniform over [1, 1000] Mpc, and the prior for $\theta_{\text{JN}}$ is sinusoidal from 0 to $\pi$. With a Gaussian likelihood over the parameter space, we use the \texttt{nestle} sampler in Bilby to estimate $d_L$ and $\theta_{\text{JN}}$. The $d_L$ samples are obtained for 10 sources each from the three $\langle M_{\star} \rangle$ populations of BNS models for the three GW detector configurations \texttt{LVK}, \texttt{LVKI}, and \texttt{CE+ET}. 

\textit{Parameter estimation using Gaussian method: } We estimate the median $\sigma_{d_L}/d_L$ for \texttt{LVK}, \texttt{LVKI}, and \texttt{CE+ET} from Bilby. Then we use that median error to model the Gaussian posterior on the luminosity distance for all the three cases of $\langle M_{\star} \rangle$. 
For \texttt{LVK}, \texttt{LVKI}, and \texttt{CE+ET} configuration, we consider the value of $\sigma_{d_L}/d_L$ as $19\%$, $15.5\%$ and $1.6\%$ respectively. We generate $d_L$ samples, for the entirety of the host populations from a Gaussian distribution with the mean obtained from eq. \ref{lum_dis_eqn} and the percentage standard deviation obtained from the median values for every detector configuration. We then execute the $v_p$ correction formalism for all host galaxies each across three different detector configurations.

\section{Impact of peculiar velocity on Hubble constant estimation}\label{section:peculiar velocity and hubble constant}

As discussed in section \ref{section:motivation}, the peculiar motion of a galaxy contaminates the redshift estimate which affects the distance-redshift relation resulting in a biased inference of $H_0$. Hence, \citet{Mukherjee_2021} developed a Bayesian formalism to incorporate and correct for the peculiar velocity contamination for inferring the Hubble constant from bright standard sirens. The equation for the framework is given by,
\begin{equation}\label{eq:framework}
    \begin{split}
        P\left(H_0|\left\{D_{GW}\right\},\left\{\hat{z}\right\}\right)\propto\prod_{n=\mathbf{1}}^{N_{\rm GW}}\int{dd_L^ndv_p^n}\mathbf{\mathcal{L}}\left(d_L^n|H_0,v_p^n,z_n,\widehat{u}_n,D_{GW}^n\right)\\ P\left(z_n|\widehat{u}_n\right) P\left(v_p^n|M,\widehat{u}_n\right) \Pi (H_0),
    \end{split}
\end{equation}
where $P\left(H_0|\left\{D_{GW}\right\},\left\{\hat{z}\right\}\right)$ is the posterior PDF of $H_0$ given the GW data for $N_{GW}$ sources and their observed redshifts estimated from the EM counterpart; $\mathcal{L}$ is the likelihood on the luminosity distance $d_L$ which is assumed to be Gaussian; $P(z|\hat{u})$ is posterior of redshift estimate at source; $P(v_p|M,\hat{u})$ is the posterior of the peculiar velocity of the host galaxy that has a halo of mass $M$ and located at sky position $\hat{u}$ (RA, Dec) ; and $\Pi$($H_0$) is the prior on the value of $H_0$.

Having three populations of galaxies (discussed in section \ref{subsection:GLADE+}) from GLADE+, we generate samples of $v_p$ for each galaxy from a normal distribution with standard deviation $\sigma_{v_p}$. This forms the peculiar velocity posterior ($P(v_p|M,\hat{u})$). We then use the $d_L$ posteriors obtained using (i) luminosity distance posterior from Bilby and (ii) Gaussian luminosity distance posterior to obtain posterior distribution on Hubble constant $H_0$ from the mock samples.

\subsection{$H_0$ correction for GW sources with PE from Bilby}
To implement the $v_p$ correction (eq. \ref{eq:framework}), we use a flat prior on $H_0$ over the range [20, 150] km s$^{-1}$ Mpc$^{-1}$, and with the $d_L$ samples obtained using Bilby, we use the ensemble sampler from the \texttt{emcee} \citep{emcee_paper} package to implement the Metropolis-Hastings algorithm. The code samples from the above defined prior on $H_0$ and evaluate the likelihood by calculating the corresponding model value for $d_L$ using the $v_p$ realizations. After marginalizing over the $v_p$ uncertainties, we get the corresponding $H_0$ samples and obtain the posterior probability distribution function for $H_0$ for individual sources using kernel smoothing with a scale of $\sim 0.013$ km s$^{-1}$ Mpc$^{-1}$. We then normalize the 10 individual posteriors and combine them to obtain a combined $H_0$ posterior for each $\langle M_{\star} \rangle$ population for each detector configuration.

In Fig.  \ref{fig:H0_posteriors} we present the combined posteriors for 10 sources from the three $\langle M_{\star} \rangle$ cases for \texttt{LVK}, \texttt{LVKI} and \texttt{CE+ET} detector configurations respectively. For the \texttt{LVK} configuration, the joint posteriors for the three populations have a similar spread. With the addition of the LIGO-India detector, as the uncertainty on the estimation of luminosity distance decreases, we start to see the impact of the stellar mass of the host galaxy on the precision of $H_0$. This effect becomes more evident from the \texttt{CE+ET} configuration as it has the most precise $d_L$ estimation of the three.

To elucidate the role of distance uncertainty, we plot the precision on $H_0$ as a function of the log of the expectation value of the host stellar mass for the three detector networks as shown in Fig.  \ref{fig:error_stellar_mass_bilby}. We see that with the \texttt{CE+ET} detectors, for the $\langle M_{\star} \rangle = 10^9 M_{\odot}, 10^{10} M_{\odot}$ and $10^{11} M_{\odot}$ populations, the respective $H_0$ precisions are $\sim 1.1\%, 1.9\%$ and $2.3\%$. We observe that for \texttt{CE+ET}, as we go from low-mass BNS host galaxies to high-mass BNS host galaxies, the precision on $H_0$ inferred from a BNS merger in the host drops by a factor of $\sim 2$.

Fig.  \ref{fig:combined glade_error 10 to 50} shows the precision on $H_0$ as a function of the number of bright siren events cumulatively combined for the three detector configurations, denoted by three different markers. For each marker, the red, yellow, and blue lines represent the $\langle M_{\star} \rangle = 10^9$ M$_\odot$, $10^{10}$ M$_\odot$, $10^{11}$ M$_\odot$ populations respectively. Here we see the interplay between the distance error and the peculiar velocity bias. With the \texttt{CE+ET} configuration, the contribution from the peculiar velocity uncertainty to the precision on $H_0$ far exceeds that from the distance uncertainty. Hence we clearly see the impact of the host properties in terms of the separation between the three dashed lines with lozenge ($\blacklozenge$) markers. As the $\langle M_{\star} \rangle = 10^{11}$ M$_\odot$ population contains extremely heavy galaxies of
the order of $10^{11}$ M$_\odot$ -$10^{12}$ M$_{\odot}$, the peculiar velocity contribution from the host galaxies in this population is much greater than that from the other two populations. As a result, the blue dashed line gets flattened at $N_{GW}=3$ and the precision on $H_0$ does not improve even after combining 10 sources.

\begin{figure}
    \centering
    \includegraphics[width=0.48\textwidth]{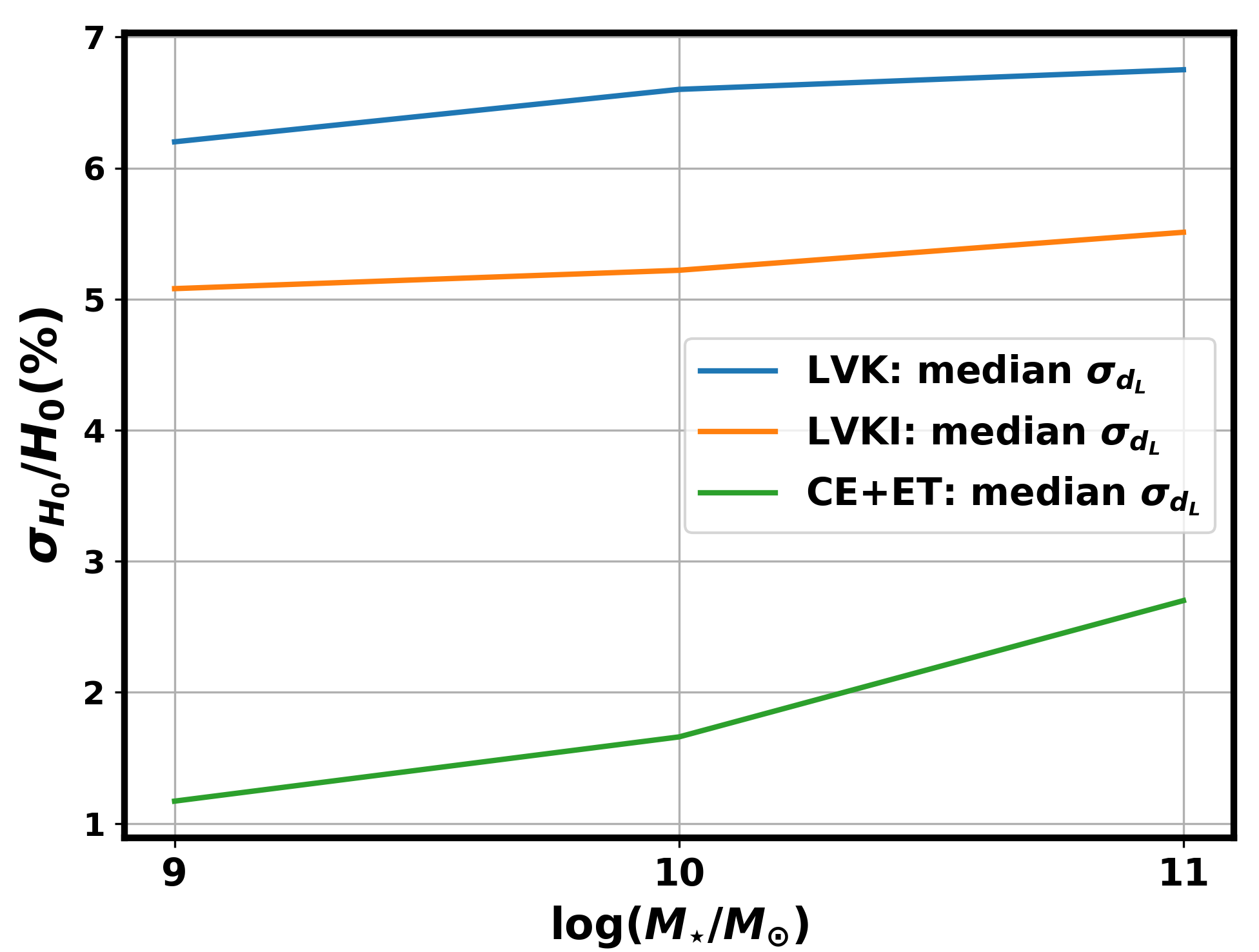}
    \caption{The precision on $H_0$ as a function of the log-stellar mass for the three detector configurations for the exact same sources as in Fig.  \ref{fig:error_stellar_mass_bilby}. The luminosity distance samples are obtained from a Gaussian distribution with a standard deviation obtained from the median of the percent standard deviation for 30 sources (10 per population) per detector configuration.}
    \label{fig:error_stellar_mass_gaussian}
\end{figure}

\begin{figure*}
    \centering
    \includegraphics[width=\textwidth]{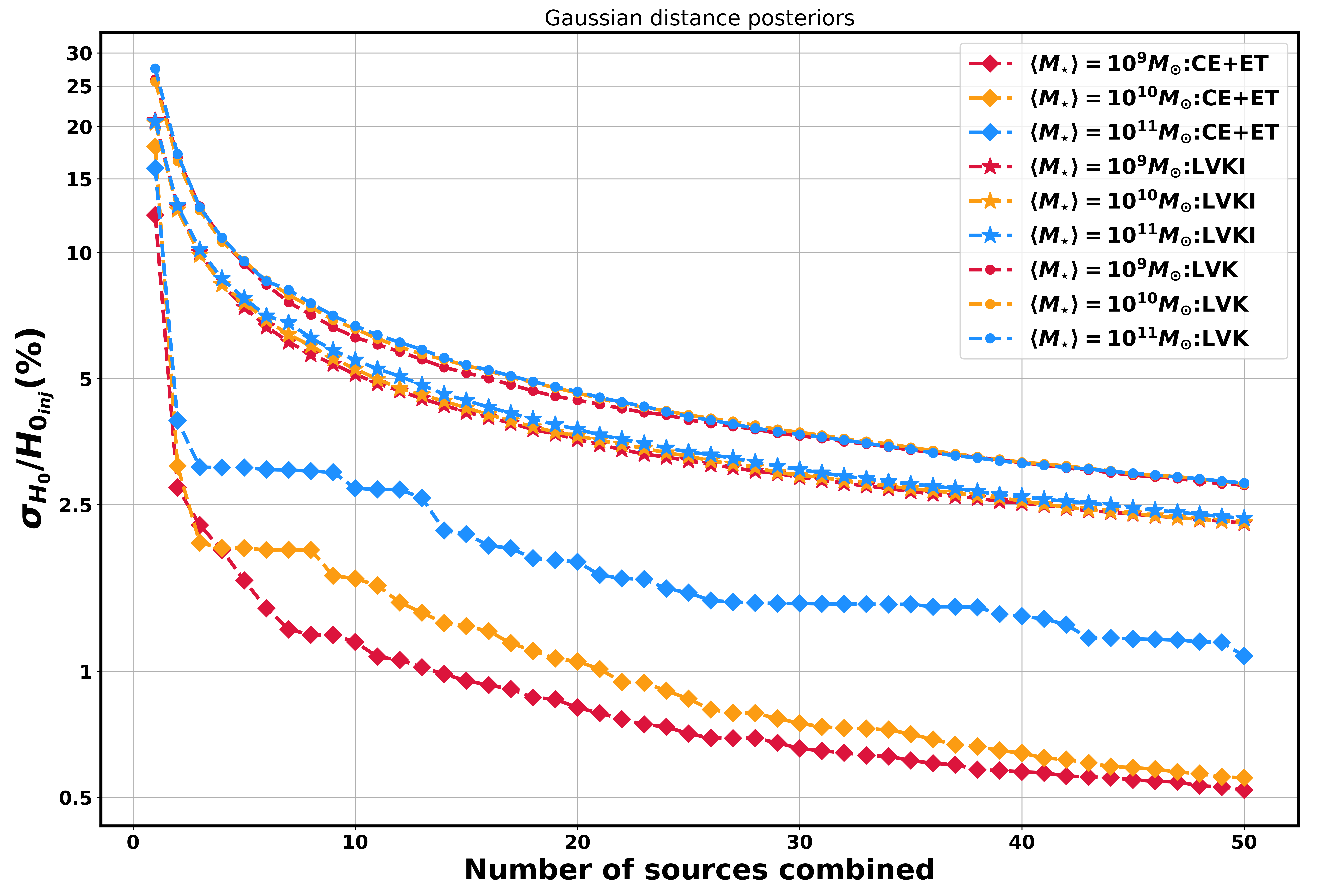}
    \caption{Precision on $H_0$ as a function of the number of sources cumulatively combined for the three detector configurations (denoted by distinct markers) for the three $\langle M_{\star} \rangle$ populations (distinctly colored). The distance posteriors are approximated with a Gaussian with percent standard deviation obtained from the median value (refer to section \ref{subsection: Gaussian_approximation_distance}) for 10 sources for each detector configuration and each case.}
    \label{fig:combined_det}
\end{figure*}

\subsection{$H_0$ correction for GW sources with PE from Gaussian approximation}\label{subsection: Gaussian_approximation_distance}
Before we perform an analysis of the entire chosen population of BNS host galaxies, we first compare the results from the non-Gaussian $d_L$ posteriors with Gaussian ones. For the exact same sources, we make a plot, similar to Fig.  \ref{fig:error_stellar_mass_bilby} in Fig.  \ref{fig:error_stellar_mass_gaussian}. Comparing the similar colored lines from the two figures, we observe that with the Gaussian approximation with median percent standard deviation, the precision of $H_0$ for \texttt{LVK} and \texttt{LVKI} configurations is slightly poorer than that for the non-Gaussian $d_L$ posteriors. For the \texttt{CE+ET} configuration, the Gaussian case shows a steeper slope from $\langle M_{\star} \rangle = 10^{10} M_{\odot}$ to $10^{11} M_{\odot}$. This assures that with the median $\sigma_{d_L}/d_L$, we are not underestimating the contribution from the distance uncertainty to the precision on $H_0$.

Fig.  \ref{fig:combined_det} shows the precision on $H_0$ as a function of the number of bright siren events cumulatively combined (similar to Fig.  \ref{fig:combined glade_error 10 to 50} but with 50 sources), with Gaussian approximation for the luminosity distance with median $\sigma_{d_L}/d_L$ for the three detector configurations. The three detector configurations are denoted by three different markers, similar to Fig.  \ref{fig:combined glade_error 10 to 50}. For each marker, the red, yellow, and blue lines represent the $\langle M_{\star} \rangle = 10^9 M_{\odot}, 10^{10} M_{\odot}, 10^{11} M_{\odot}$ populations respectively. It can be seen that for the \texttt{CE+ET} configuration, for BNS host galaxies with a stellar mass of the order of $10^9$ M$_{\odot}$, a one percent precision on $H_0$ can be attained with $\approx 15$ bright siren events. If the host stellar mass is of the order 10$^{10}$ M$_{\odot}$, we need more than 20 events, and for 10$^{11}$ M$_{\odot}$, we need at least 50 bright siren events to reach the 1\%-precision on $H_0$. Similar to Fig.  \ref{fig:combined glade_error 10 to 50}, we see that the blue lozenges in Fig.  \ref{fig:combined_det}, representing the \texttt{CE+ET} configuration for the $\langle M_{\star} \rangle = 10^{11} M_{\odot}$ population, reach saturation and tend to get flattened as we combine more and more posteriors. As the peculiar velocity uncertainty far exceeds the distance error (1.6\%), combining more posteriors accounts for the distance error and highlights the contribution from the host galaxy properties.

\section{Conclusion and future outlook} \label{section:conclusion}
Binary neutron stars (BNS), being a class of bright standard sirens, are the ideal candidates to resolve the Hubble tension. However, since they are detected at low redshifts, the peculiar motions of their host galaxies contaminate the redshift and contribute significantly to the error budget on the inference of the Hubble constant ($H_0$). Hence to correct for the peculiar velocity contamination, it is necessary to estimate the peculiar velocity accurately. The peculiar velocity estimates are driven by the properties of the host galaxies of GW sources. Along with appropriate modeling of the source population, to discern the role of their host properties, we need to also analyze the impact of the distance uncertainty of a GW source on the $H_0$ estimation. Hence we considered three different ground-based detector configurations. These three configurations each lead to a different approximate uncertainty on the distance inferred from them.

With different detectors, we get different accuracy in the estimation of the luminosity distance of the GW sources. First, we consider the LIGO (Hanford and Livingston) - Virgo - KAGRA (\texttt{LVK}) network. We see that with \texttt{LVK}, the median uncertainty on the $d_L$ estimate for 30 sources is $19\%$. If the $d_L$ uncertainty exceeds $\sim 20\%$, the impact of $v_p$ uncertainty is shrouded and $\sigma^2_{d_L}$ starts leading $\sigma^2_{H_0}$. $\sigma^2_{d_L}$ is largely dependent on the ``loudness" of the GW chirp which depends on the intrinsic properties of the GW sources.

The signal-to-noise ratio (SNR) of a GW incident on a given detector can vary depending on the number of detectors. Hence to achieve greater accuracy on $d_L$, we consider another detector (LIGO-India) along with \texttt{LVK}. We observe that with the addition of the LIGO-India detector, we get a much better constraint on the $d_L$ of the GW sources. The median uncertainty on the $d_L$ estimate for the same 30 sources is $15.5\%$. The improved $d_L$ estimates illuminate the contribution of $v_p$ uncertainty and thus make way for the GW host property dependence to show up in the form of a significant difference in the $H_0$ precision for the BNS host populations.

To minimize the dominance of the error in the distance measurement, we considered a third configuration involving the Cosmic Explorer and Einstein Telescope (\texttt{CE+ET}) detectors. The \texttt{CE+ET} configuration has a median distance uncertainty of $1.6\%$ which helps explore the host property dependence of peculiar velocity to its full potential.

The results from this study present a forecast on the impact of the host properties of GW sources, particularly BNSs on the peculiar velocity estimation and subsequently on the inference of $H_0$. The results for these real galaxies from the GLADE+ catalog show that the stellar mass has a major impact on the peculiar velocity estimation and as a consequence, it significantly affects the inference of $H_0$. To shed light on the impact of the host properties on peculiar velocity estimation, we need to minimize the dominance of the distance uncertainty. We saw that a BNS merger event detected by \texttt{CE+ET} configuration located in a host galaxy with stellar mass $10^9 M_{\odot}$ yields a more precise estimate of $H_0$ by a factor of 2 than a host with stellar mass $10^{11} M_{\odot}$. This implies that host galaxies with more stellar mass will have a larger uncertainty on their peculiar velocity. The key takeaway is that it is equally important to measure the luminosity distance with greater accuracy as it is important to correct for the peculiar velocity of the GW source, to achieve precision measurement of the Hubble constant.

In conclusion, the variance on $H_0$ is an interplay between the luminosity distance uncertainties and the peculiar velocity uncertainties. Hence for a precise estimation of $H_0$, it is necessary to accurately determine distances to GW sources while simultaneously estimating their peculiar velocities with precision. Lastly, we saw that by combining more sources, typically of the order of 50, we can reach a $1\%$ precision on $H_0$. However, this will depend on the host properties of the GW sources due to their peculiar velocity contamination. We saw that with \texttt{CE+ET} configuration, BNS mergers from less massive host galaxies will lead to a precision of 1\% on $H_0$ with approximately 15 events as compared to 40-50 events for high mass galaxies. In the future, with accurate modeling of peculiar velocity and accurately mitigating its contamination, one can reach a  $1\%$ measurement with \texttt{CE+ET} from 50-200 sources distributed up to $z=0.05$ depending on the underlying population of the host galaxies. From \texttt{LVK} and \texttt{LVKI} one can achieve $3-2.5\%$ measurement of the Hubble constant $H_0$ from 50 GW sources. 

\section*{Acknowledgements}
The authors thank Abhishek Sharma for reviewing the manuscript and providing useful comments as a part  of the LIGO publication and presentation policy. This work is a part of the $\langle \texttt{data|theory}\rangle$ \texttt{Universe-Lab} which is supported by the TIFR and the Department of Atomic Energy, Government of India. The authors are thankful to Mr. Parag Shah for maintaining the computer cluster of the $\langle \texttt{data|theory}\rangle$ \texttt{Universe-Lab}. The authors would like to thank the LIGO-Virgo-KAGRA Scientific Collaboration for providing the noise curves. 
This research has made use of data or software obtained from the Gravitational Wave Open Science Center (gw-openscience.org), a service of LIGO Laboratory, the LIGO Scientific Collaboration, the Virgo Collaboration, and KAGRA. LIGO Laboratory and Advanced LIGO are funded by the United States National Science Foundation (NSF) as well as the Science and Technology Facilities Council (STFC) of the United Kingdom, the Max-Planck-Society (MPS), and the State of Niedersachsen/Germany for support of the construction of Advanced LIGO and construction and operation of the GEO600 detector. Additional support for Advanced LIGO was provided by the Australian Research Council. Virgo is funded, through the European Gravitational Observatory (EGO), by the French Centre National de Recherche Scientifique (CNRS), the Italian Istituto Nazionale di Fisica Nucleare (INFN) and the Dutch Nikhef, with contributions by institutions from Belgium, Germany, Greece, Hungary, Ireland, Japan, Monaco, Poland, Portugal, Spain. The construction and operation of KAGRA are funded by Ministry of Education, Culture, Sports, Science and Technology (MEXT), and Japan Society for the Promotion of Science (JSPS), National Research Foundation (NRF) and Ministry of Science and ICT (MSIT) in Korea, Academia Sinica (AS) and the Ministry of Science and Technology (MoST) in Taiwan. This material is based upon work supported by NSF's LIGO Laboratory which is a major facility fully funded by the National Science
Foundation. We acknowledge the use of the following packages in this work: Astropy \citep{astropy:2013,astropy:2018,astropy:2022}, Bilby \citep{Ashton_2019,lalsuite,swiglal}, emcee: MCMC Hammer \citep{emcee_paper}, Matplotlib \citep{Matplotlib:2007}, NumPy \citep{numpy}, Pandas \citep{pandas}, SciPy \citep{SciPy}, Seaborn \citep{Seaborn}.

 \section*{Data Availability}
The data underlying this article will be shared at the request to the corresponding author. 

\bsp	
\bibliographystyle{mnras}
\bibliography{arXiv-submission/version_2}
\nocite{*}
\label{lastpage}
\end{document}